\documentstyle[prl,aps,twocolumn,epsf]{revtex}
\def\break#1{\pagebreak \vspace*{#1}}
\begin{document}

\draft

\title{Ermakov approach for empty FRW minisuperspace oscillators}

\author{Haret C. Rosu
}

\address{
{Instituto de F\'{\i}sica de la Universidad de Guanajuato, Apdo Postal
E-143, Le\'on, Gto, M\'exico}\\
}

\maketitle
\widetext

\begin{abstract}

The formal Ermakov approach for empty
FRW minisuperspace cosmological models of arbitrary Hartle-Hawking factor
ordering and the corresponding squeezing features are briefly discussed as
a possible means of describing cosmological evolution.

\end{abstract}
\vskip 0.2in


\narrowtext


The introduction of Ermakov-type (adiabatic) invariants \cite{erm}
may prove a useful method of investigating evolutionary and chaotic dynamics
problems in the ``quantum" cosmological framework \cite{work}.
Moreover, the method
of adiabatic invariants is intimately related to geometrical angles and
phases \cite{book}, so that one may think of cosmological
Hannay's angles as well as various types of topological phases
as those of Berry and Pancharatnam \cite{dutta}.

My purpose in the following is to apply the formal Ermakov scheme to the
simplest cosmological oscillators, namely the empty
Friedmann-Robertson-Walker (FRW) ``quantum" universes.
The Wheeler-DeWitt (WDW)
minisuperspace equation for that case can be written down as
follows 
\begin{equation} \label{1}
\frac{d^2\Psi}{d\Omega^2}+Q\frac{d\Psi}{d\Omega}-e^{-4\Omega}\Psi (x) =0~,
\end{equation}
where $Q$ is the Hartle-Hawking parameter for the factor ordering \cite{hh}
that is kept as a free parameter,
and $\Omega$ is Misner's time \cite{mi}.

Eq.~(1) can be mapped in a known way \cite{clas}
to the canonical equations for a classical point
particle of 
mass $M=e^{Q\Omega}$ and taking $\Psi =q$ as a generalized coordinate
and $\Psi ^{'}$ as the momentum, leading to
\begin{eqnarray}
\frac{dq}{d\Omega}&=&e^{-Q\Omega}p~\\
\frac{dp}{d\Omega}&=&e^{(Q-4)\Omega}q~.
\end{eqnarray}
These equations describe the canonical
motion for a classical point universe as derived from the
time-dependent Hamiltonian (of the inverted oscillator type \cite{inv})
\begin{equation} \label{4}
H(\Omega)=e^{-Q\Omega}\frac{p^2}{2}-e^{(Q-4)\Omega}\frac{q^2}{2}~.
\end{equation}
For this FRW Hamiltonian the triplet of phase-space
functions $T_1=\frac{p^2}{2}$, $T_2=pq$,
and $T_3=\frac{q^2}{2}$ forms a dynamical Lie algebra (i.e.,
$H=\sum _{n}h_{n}(\Omega)T_{n}(p,q)$) which is closed with
respect to the Poisson bracket, or more exactly
$\{T_1,T_2\}=-2T_1$, $\{T_2,T_3\}=-2T_3$, $\{T_1,T_3\}=-T_2$. The FRW
Hamiltonian can be written down as $H=e^{-Q\Omega}T_1-e^{(Q-4)\Omega}T_3$.
\break{1.178in}
The Ermakov invariant $I$ belongs to the dynamical algebra
\begin{equation} \label{5}
I=\sum _{r}\epsilon _{r}(\Omega)T_{r}~,
\end{equation}
and by means of

\begin{equation} \label{6}
\frac{\partial I}{\partial \Omega}=-\{I,H\}~,
\end{equation}
one is led to the following equations for the unknown functions
$\epsilon _{r}(\Omega)$
\begin{equation} \label{7}
\dot{\epsilon} _{r}+\sum _{n}\Bigg[\sum _{m}C_{nm}^{r}h_{m}
(\Omega)\Bigg]\epsilon _{n}=0~,
\end{equation}
where $C_{nm}^{r}$ are the structure constants of the Lie algebra that
have been already given above. Thus, we get
\begin{eqnarray} \nonumber
\dot{\epsilon} _1&=&-2e^{-Q\Omega}\epsilon _2 \\
\dot{\epsilon} _2&=&-e^{(Q-4)\Omega}\epsilon _1-e^{-Q\Omega}\epsilon _3\\
\dot{\epsilon} _3&=&-2e^{(Q-4)\Omega}\epsilon _2~.    \nonumber
\end{eqnarray}
The solution of this system can be readily obtained by setting
$\epsilon _1=\rho ^2$ giving $\epsilon _2=-e^{Q\Omega}\rho \dot{\rho}$ and
$\epsilon _3=e^{2Q\Omega}\dot{\rho} ^2 +
\frac{1}{\rho ^2}$, where $\rho$ is the solution of the Milne-Pinney equation
\cite{mp}
\begin{equation} \label{9}
\ddot{\rho}+Q\dot{\rho}-e^{-4\Omega}\rho=\frac{e^{-2Q\Omega}}{\rho ^3}~.
\end{equation}
In terms of the function $\rho (\Omega)$ the Ermakov invariant can be
written as follows \cite{l}
\begin{equation} \label{10}
I=\frac{(\rho p-e^{Q\Omega}\dot{\rho}q)^2}{2}+\frac{q^2}{2\rho ^2}~.
\end{equation}
Next, we calculate the time-dependent generating function allowing one to
pass to new canonical variables for which $I$ is chosen to be the
new ``momentum"
\begin{equation} \label{11}
S(q,I,\vec{\epsilon}(\Omega))=\int ^{q}dq^{'}p(q^{'},I,\vec{\epsilon}
(\Omega))~,
\end{equation}
leading to
\begin{eqnarray}   \nonumber         
S(q,I,\vec{\epsilon}
(\Omega))&=&e^{Q\Omega}\frac{q^2}{2}\frac{\dot{\rho}}{\rho}+
I{\rm arctan}\Bigg[\frac{q}{\sqrt{2I\rho ^2-q^2}}\Bigg]+\\
         &+&\frac{q\sqrt{2I\rho ^2-q^2}}{2\rho ^2}~, 
\end{eqnarray}
where we have put to zero the constant of integration.
Thus
\begin{equation} \label{13}
\theta=\frac{\partial S}{\partial I}={\rm arctan}
\Big(\frac{q}{\sqrt{2I\rho ^2-q^2}}\Big)~.
\end{equation}
Moreover, the canonical variables are now
\begin{equation} \label{14}
q=\rho \sqrt{2I}\sin \theta ~,
\end{equation}
and
\begin{equation}  \label{15}
p=\frac{\sqrt{2I}}{\rho}\Big(\cos \theta+
e^{Q\Omega}\dot{\rho}\rho\sin \theta\Big)~.
\end{equation}
The dynamical angle will be
\begin{equation} \label{16}
\Delta \theta ^{d}=
\int _{\Omega _{0}}^{\Omega}
\langle\frac{\partial H_{\rm{new}}}{\partial I}\rangle
d\Omega ^{'}=
\int _{0}^{\Omega}\Bigg[\frac{e^{-Q\Omega '}}{\rho ^2}-\frac{\rho ^2}{2}
\frac{d}{d\Omega ^{'}}\Big(\frac{\dot{\rho}}{\rho}\Big)\Bigg]d\Omega ^{'}~,
\end{equation}
whereas the geometrical angle reads
\begin{equation}  \label{17}
\Delta \theta ^{g}=\frac{1}{2}\int _{\Omega _0}^{\Omega}
\Bigg[\frac{d}{d\Omega ^{'}}
(e^{Q\Omega ^{'}}\dot{\rho}\rho)-2e^{Q\Omega ^{'}}\dot{\rho}^2\Bigg]
d\Omega ^{'}~.
\end{equation}

The total change of angle will be
\begin{equation}  \label{17b}
\Delta \theta =\int _{\Omega _{0}}^{\Omega}\frac{e^{-Q\Omega ^{'}}}{\rho ^2}
d\Omega ^{'}~.
\end{equation}
On the Misner time axis, going to $-\infty$ means going to the origin of the
universe, whereas $\Omega _{0}=0$ means the present epoch. Thus, using
these cosmological limits in Eq.~(18) one can see that the total change
of angle can be written as the Laplace transform (up to a sign)
of the inverse square of the Milne-Pinney function, $\Delta \theta=
-L_{1/\rho ^{2}}(Q)$.

Passing now to the quantum Ermakov problem we turn $q$
and $p$ into quantum-mechanical operators, $\hat{q}$ and
$\hat{p}=-i\hbar\frac{\partial}{\partial q}$, but keeping the auxiliary
function $\rho$ as a $c$ number. The Ermakov invariant is a
constant Hermitian operator (if one works with real Milne-Pinney functions)
\begin{equation}  \label{18}
\frac{dI}{d\Omega}=\frac{\partial I}{\partial \Omega}+\frac{1}{i\hbar}[I,H]=0
\end{equation}
and the FRW time-dependent Schr\"odinger equation reads
\begin{equation}  \label{19}
i\hbar\frac{\partial}{\partial \Omega}|\psi (q,\Omega)\rangle=
H(\Omega)|\psi(q,\Omega)\rangle ~.
\end{equation}

The goal now is to find the eigenvalues of $I$
\begin{equation}  \label{20}
I|\phi _{n}(q,\Omega)\rangle=\kappa _{n}|\phi _{n}(q,\Omega)\rangle
\end{equation}
and to write the explicit superposition form of the general solution of
Eq.~(20) \cite{lr}
\begin{equation}  \label{21}
\psi(q,\Omega)=\sum _{n}C_{n}e^{i\alpha _{n}(\Omega)}\phi _{n}(q,\Omega)~,
\end{equation}
where $C_{n}$ are superposition constants, $\phi _{n}$ are the (orthonormal)
eigenfunctions of $I$, and the phases $\alpha _{n}(\Omega)$,
(so-called Lewis phases) 
are to be found from the equation
\begin{equation}  \label{22}
\hbar \frac{d\alpha _{n}(\Omega)}{d\Omega}=\langle \psi _{n}|i\hbar
\frac{\partial}{\partial \Omega}-H|\psi _{n}\rangle~.
\end{equation}
The key point for the quantum Ermakov problem is to perform a clever unitary
transformation in order to obtain a transformed
eigenvalue problem for the new Ermakov invariant $I^{'}=UIU^{\dagger}$
possessing time-independent eigenvalues.
It is easy to get the required unitary transformation
as $U=\exp [-\frac{i}{\hbar}(e^{Q\Omega})
\frac{\dot{\rho}}{\rho}\frac{\hat{q}^2}{2}]$ and
the new Ermakov invariant will be $I^{'}=\frac{\rho ^{2}\hat{p}^2}{2}+
\frac{\hat{q}^{2}}{2\rho ^{2}}$. Therefore, its eigenfunctions are
$\propto e^{-\frac{\theta ^2}{2\hbar}}H_{n}(\theta/\sqrt{\hbar})$, where
$H_{n}$ are Hermite polynomials, $\theta=\frac{q}{\rho}$, and the eigenvalues
are $\kappa _{n}=\hbar(n+\frac{1}{2})$.
Thus, one can write the eigenfunctions $\psi _{n}$ as follows
\begin{equation}  \label{23}
\psi _{n}\propto \frac{1}{\rho ^{\frac{1}{2}}}
\exp \Big(\frac{1}{2}\frac{i}{\hbar}(e^{Q\Omega})
\frac{\dot{\rho}}{\rho}q^2\Big)\exp\Big(-\frac{q^2}{2\hbar \rho ^2}\Big)
H_{n}\Big(\frac{1}{\sqrt{\hbar}}\frac{q}{\rho}\Big)~.
\end{equation}
The factor $1/\rho ^{1/2}$ has been introduced for normalization reasons.
Using these functions and performing
simple calculations one is lead to the geometrical phase
\begin{equation}  \label{23b}
\alpha _{n}^{g}=-\frac{1}{2}\Big(n+\frac{1}{2}\Big)\int _{\Omega _0}^{\Omega}
\Bigg[(\ddot{\rho}\rho)-\dot{\rho}^2\Bigg] d\Omega ^{'}~.
\end{equation}
For the general phases entering the Ermakov superposition in Eq.~(22)
one gets the following result
\begin{equation}  \label{24}
\alpha _{n}(\Omega)=-(n+\frac{1}{2})\int _{\Omega _{0}}^{\Omega}
\frac{e^{-Q\Omega '}d\Omega '}{\rho ^{2}}
\end{equation}
and in the cosmological limits one finds once again a Laplace transform of the
inverse square of the Milne-Pinney function,
$\alpha _{n}=(n+\frac{1}{2})L_{1/\rho ^{2}}(Q)$.

The Ermakov procedure allows to construct cosmological squeezed states in a
very convenient way \cite{sq1,sq2}.  For this one makes use of the
factorization of the Ermakov invariant $I=\hbar(bb^{\dagger}+\frac{1}{2})$,
where
\begin{eqnarray} 
b          &=&(2\hbar \rho)^{-1/2}
  [\rho ^{-1/2}q+i\rho ^{1/2}(\rho p-e^{Q\Omega}\dot{\rho}q)]\\
b^{\dagger}&=&
(2\hbar \rho)^{-1/2}[\rho ^{-1/2}q-i\rho ^{1/2}(\rho p-e^{Q\Omega}
\dot{\rho}q)]~.
\end{eqnarray}
Let us now consider a reference Misner-time-independent oscillator with
the Misner frequency fixed at an arbitrary epoch $\Omega _{0}$ for which
one can write the common factoring operators
\begin{eqnarray}  
a          &=&(2\hbar \omega _{0})^{-1/2}[\omega _{0}q+ip]\\
a^{\dagger}&=&(2\hbar \omega _{0})^{-1/2}[\omega _{0}q-ip]~.
\end{eqnarray}
The connection between the $a$ and $b$ pairs is given by
\begin{eqnarray} 
b(\Omega)          &=&\mu(\Omega)a+\nu(\Omega)a^{\dagger}\\
b^{\dagger}(\Omega)&=&\mu ^{*}(\Omega)a^{\dagger}+\nu ^{*}
(\Omega)a^{\dagger}~, 
\end{eqnarray}
where
\begin{equation} \label{sq1}
\mu(\Omega)=(4\omega _{0})^{-1/2}[\rho ^{-1}-ie^{Q\Omega}
\dot{\rho}+\omega _{0}\rho]
\end{equation}
and
\begin{equation} \label{sq2}
\nu(\Omega)=(4\omega _{0})^{-1/2}[\rho ^{-1}-ie^{Q\Omega}
\dot{\rho}-\omega _{0}\rho]
\end{equation}
fulfill the well-known relationship $|\mu(\Omega)|^{2}-|\nu(\Omega)|^{2}=1$.
The corresponding uncertainties are known to be
$(\Delta q)^{2}=\frac{\hbar}{2\omega _{0}}|\mu - \nu|^{2}$,
$(\Delta p)^{2}=\frac{\hbar \omega _{0}}{2}|\mu + \nu|^{2}$, and
$(\Delta q)(\Delta p)= \frac{\hbar}{2}|\mu +\nu||\mu -\nu|$ showing that
in general the Ermakov squeezed states are not minimum uncertainty states
\cite{sq2}.

In conclusion, a simple cosmological application of the (classical
and quantum) Ermakov procedure has been presented on the base of a
classical point particle representation of the FRW WDW equation.
It is also to be noted that the Ermakov invariant
is equivalent to the Courant-Snyder one \cite{cs},
allowing thus in a certain sense a beam physics approach to
cosmological evolution.

\section*{ Acknowledgment}

This work was partially supported by CONACyT Project 458100-5-25844E.



\begin{thebibliography} {99}


\bibitem{erm}
         Ermakov V P 1880 {\em Univ. Izv. Kiev. Ser. III} {\bf 9} 1

\bibitem{work}
         See for example,
         Cotsakis S, Lemmer R L, and Leach P G L 1998 {\em Phys. Rev.} D
         {\bf 57} 4691;
         Kim S P 1996 {\em Class. Quantum Grav.} {\bf 13} 1377

\bibitem{book}
         Berry M V 1984 {\em Proc. R. Soc. London} A {\bf 392} 45;
         Hannay J H 1985 {\em J. Phys.} A {\bf 18} 221;
         Shapere A and Wilczek F {\em Geometric Phases in Physics}
         (World Scientific, Singapore, 1989); Anandan J, Christian J,
         and Wanelik K 1997 {\em Am. J. Phys.} {\bf 65} 180

\bibitem{dutta}
         Dutta D P 1993 {\em Phys. Rev.} D {\bf 48} 5746;
         1993 {\em Mod. Phys. Lett.}
         A {\bf 8} 191 and 601


\bibitem{hh}
         Hartle J and Hawking S W 1983 {\em Phys. Rev.} D {\bf 28} 2960

\bibitem{mi}
         Misner C W 1969 {\em Phys. Rev. Lett.} {\bf 22} 1071;
         1969 {\em Phys. Rev.} {\bf 186} 1319 and 1328

\bibitem{clas}
         For a recent paper see Mielnik B and Reyes M A 1996
         {\em J. Phys.} A {\bf 29} 6009

\bibitem{inv}
         See for example, Baskoutas S {\em et al.} 1993
         {\em J. Phys.} A {\bf 26} L819






\bibitem{mp}
         Milne W E 1930 {\em Phys. Rev.} {\bf 35} 863;
         Pinney E 1950 {\em Proc. Am. Math. Soc.} {\bf 1} 681

\bibitem{l}
         Lewis H R Jr. 1968 {\em J. Math. Phys.} {\bf 9} 1976


         
\bibitem{lr}
         Lewis H R and Riesenfeld W B 1969 {\em J. Math. Phys.} {\bf 10} 1458





\bibitem{sq1}
         Hartley J G and Ray J R 1982 {\em Phys. Rev.} D {\bf 25} 382

\bibitem{sq2}
         Pedrosa I A 1987 {\em Phys. Rev.} D {\bf 36} 1279;
         Pedrosa I A and Bezerra V B 1997 {\em Mod. Phys. Lett.} A {\bf 12}
         1111

\bibitem{cs}
         According to the second footnote in Ref. 10.
         Courant E D and Snyder H S 1958 {\em Ann. Phys. (N.Y.)} {\bf 3} 1

\end{thebibliography}
\end{document}